\documentclass[reprint,amsmath,amssymb,aps,prstab]{revtex4-1}
\usepackage[english] {babel}
\usepackage{amsfonts,hyperref,graphicx,soul,natbib}

\usepackage{epstopdf}
\usepackage{xcolor,ulem}

\hypersetup{
  pdfauthor={S.S. Baturin}
}

\begin{document}
\title{Analytical treatment of the wakefields driven by transversely shaped beams in a planar slow-wave structure}

\author{S.S. Baturin}%
\email{s.s.baturin@gmail.com}%
\affiliation{The University of Chicago, PSD Enrico Fermi Institute, 5640 S Ellis Ave, Chicago, IL 60637, USA}%
\author{G. Andonian}%
\affiliation{Department of Physics and Astronomy, University of California, Los Angeles, California 90095-1547, USA}%
\affiliation{RadiaBeam Technologies, Santa Monica, California 90404, USA}%
\author{J.B. Rosenzweig}%
\affiliation{Department of Physics and Astronomy, University of California, Los Angeles, California 90095-1547, USA}%
\date{\today}

\begin{abstract}
The suppression of transverse wakefield effects using transversely elliptical drive beams in a planar structure is studied with a simple analytical model that unveils the geometric nature of this phenomenon.
By analyzing the suggested model we derive scaling laws for the amplitude of the longitudinal and transverse wake potentials as a function of the Gaussian beam ellipticity - $\sigma_x/a$.  
We explicitly show that in a wakefield accelerator application it is beneficial to use highly elliptical beams for mitigating transverse forces while maintaining the accelerating field. 
We consider two scaling strategies: 1) aperture scaling, where we keep a constant charge to have the same accelerating gradient as in a cylindrical structure and 2) charge scaling, where aperture is the same as in the cylindrical structure and charge is increased to match the gradient.      
\end{abstract}

\maketitle

\section{Introduction}\label{sec:intr}

Single-bunch beam breakup (BBU) effects stem from the excitation of transverse wakefields driven by off-axis particles in a particle accelerator.
One of the suggested methods of mitigating the effects of beam coupling to transverse wakefields is the introduction of a bunched beam with high transverse ellipticity in a rectangular structure with high aspect ratio \cite{FLb}.
Transverse mode control and suppression is relevant for many accelerator applications, however these effects are particularly urgent when considering advanced accelerator concepts operating at high frequency and gradient.
For example, collinear wakefield acceleration driven by intense charged particle beams in dielectric materials, has demonstrated GV/m fields \cite{Oshea:2016} in THz wakefields, and is considered a candidate method to surpass the field gradients of existing radio frequency structures.
The practicality of beam-driven wakefield acceleration for high-energy applications depends on the ability to extend the length of the acceleration process, which may be limited by BBU instability \cite{Li:2014,Qd3}.  
The beam-mode coupling can be dramatically reduced for beams with high transverse ellipticity in structures with planar geometry (consisting of two parallel planes of retarding material infinitely long in the $z$ direction, and a width in $x$ direction much larger than the beam RMS width $\sigma_x$).
Although when employing elliptical beams the longitudinal electric field behind the elliptical bunch decreases as $\sim 1/\sigma_x$, 
the deflecting force due to transverse fields scales as $\sim 1/\sigma_x^3$, leading to an advantageous scenario where the effects of the transverse forces can be heavily suppressed while maintaining substantial longitudinal fields. This effect was described using direct application of boundary conditions to Maxwell's equations for the case of a planar dielectric loaded waveguide in Ref. \cite{FLb}.
The continued studies on structures with planar geometry are relevant as these structures are commonly used in wakefield acceleration and beam phase space manipulation experiments today \cite{Antipov:2012,Andonian:2012,Gao:2018,SlacDC} and provide a natural path forward for more advanced applications.

\begin {figure*}
 \centering
\includegraphics[scale=0.25]{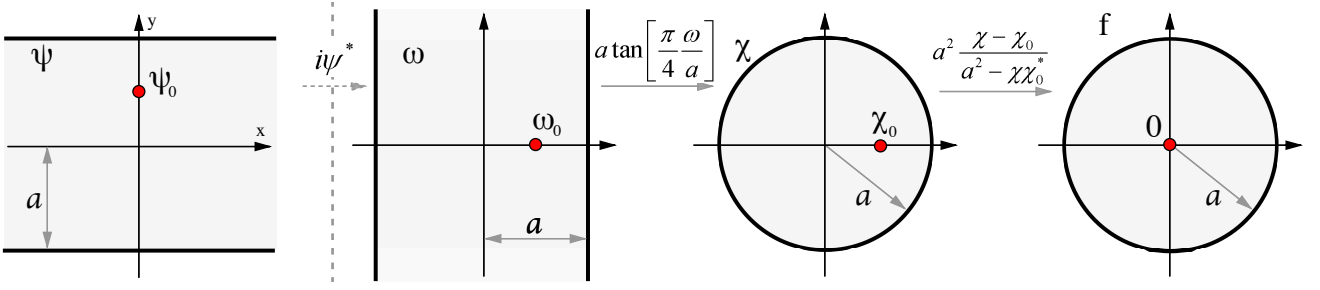}\\
\caption{Initial plane $\psi$ and schematics of the conformal mapping of the $\omega$ plane.}
\label{Fig:1}
\end {figure*}

In this paper, we further investigate the phenomenon of wakefield generation in a rectangular structure driven by an elliptical beam, by using an alternative approach based purely on geometric considerations.
We base our explanations on the limiting values of the wake potentials, previously derived in Reference~\cite{mySTAB}.
In this approach, wakefields in a longitudinally translationally invariant structure lined with layers of an arbitrary impedance  material (of dielectric, resistive, or corrugated type) have been developed. The analysis of Ref.~\cite{mySTAB} yields a derivation of expressions for the wakefields that are based on a conformal mapping method. 
It has been shown that the limiting value of losses and kicks for a point-like bunch is independent of material properties and depends only on the transverse shape of the bunch and the cross-section shape of the vacuum channel \cite{myPRL,Karl1,Karl2,BaneDc}. 
Following the conventions of Reference~\cite{mySTAB}, we consider a point-like bunch in the longitudinal coordinate with a given distribution in the transverse coordinates. 
Using this model, the transverse wake potential is calculated for various transverse beam distributions.
For charge distributions of varying ellipticity, the results are compiled to construct trade-off curves comparing the relative strengths of the longitudinal  and transverse wake potentials.
Finally, the transverse wake potentials of the elliptical beam model are directly compared to those obtained in the cylindrically symmetric case using scaled variables with respect to structure aperture and beam charge.

\section{Theoretical model of a planar structure}\label{sec:model}
The theoretical basis for our wakefield model, where the loss and kick factors are derived, has been explored in detail in Ref.\cite{mySTAB} and is summarized here.  
In this treatment, we restrict the analysis to the upper limits of the fields to obtain the relevant scaling relations for elliptical beams.
The expressions for the upper limits of the longitudinal electric field, $E_z^{0^+}$, and of the transverse component of the Lorentz force, $F_\bot$, as a function of generalized complex coordinates are 
\begin{align}
\label{eq:Ez}
 E_z^{0^+}(\omega,\omega_0)=-\frac{4 Q}{a^2}\Re[f'(\omega,\omega_0)^*f'(\omega_0,\omega_0)],
\end{align}
\begin{align}
\label{eq:Lort}
F_\bot(\omega,\omega_0,\zeta)=\frac{4 qQ \theta(\zeta) \zeta}{a^2}f''(\omega,\omega_0)^* f'(\omega_0,\omega_0).
\end{align}
Here $Q$ is the charge of the particle generating the wakefield, $q$ is the charge of the test particle, $a$ is the size of the structure aperture, $\zeta=ct-z$ is the longitudinal distance behind the particle and the test particle, and  $\theta(\zeta)$ is the Heaviside function which ensures that the field is non-vanishing only behind the particle, as dictated by causality.
The cross-section of the the structure is described as a complex plane with $\omega=x+iy$; in this analysis $F_\bot=F_x+iF_y$ and $f(\omega,\omega_0)$ is the conformal mapping function that transforms the cross-section of interest onto a circle such that the point $\omega_0$ corresponds to the center of a circle. Here the use of asterisks denotes complex conjugation. 

At this point, it should be reiterated that Eq.\eqref{eq:Ez} and Eq.\eqref{eq:Lort} represent the upper bounds for the corresponding wakefields driven by longitudinal point-like particles. The exact solutions for the complete wakefield evolution will have specific dependence on a given longitudinal bunch distribution. However, the analysis is propitious because it describes the worst-case scenario for the transverse forces as the amplitude of the point-like particle wake potential will always be greater or equal to that of a distribution for a given charge. In other words, for a charge density, $\rho_z(\zeta_0)$, the transverse force experienced by the bunch will not exceed $\int\limits_{-\infty}^\zeta F_\bot(\omega,\omega_0,\zeta-\zeta_0) \rho_z(\zeta_0)d\zeta_0$, with $F_\bot(\omega,\omega_0,\zeta-\zeta_0)$ in Eq.\eqref{eq:Lort}. In practice, optimization requires prudent augmentation of the longitudinal field with respect to the integrated transverse force.

\begin{figure*}
\includegraphics[scale=0.44]{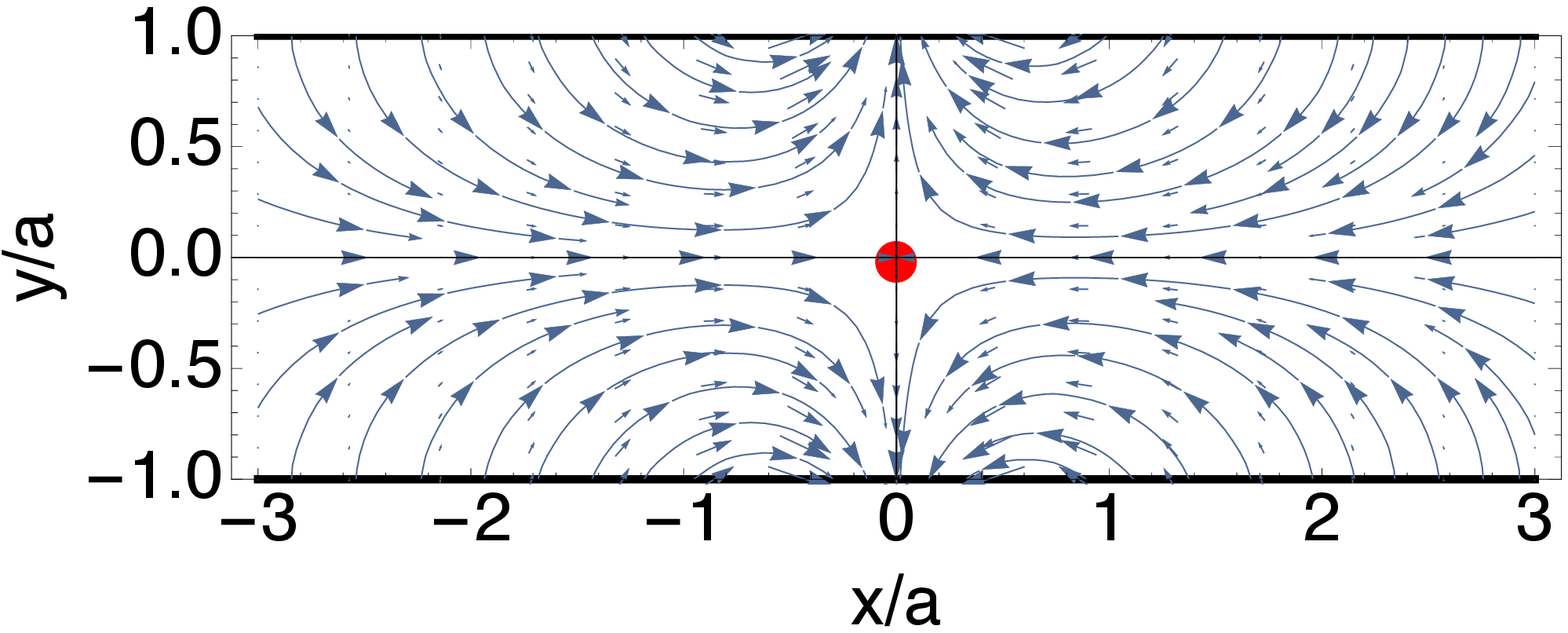}
\includegraphics[scale=0.44]{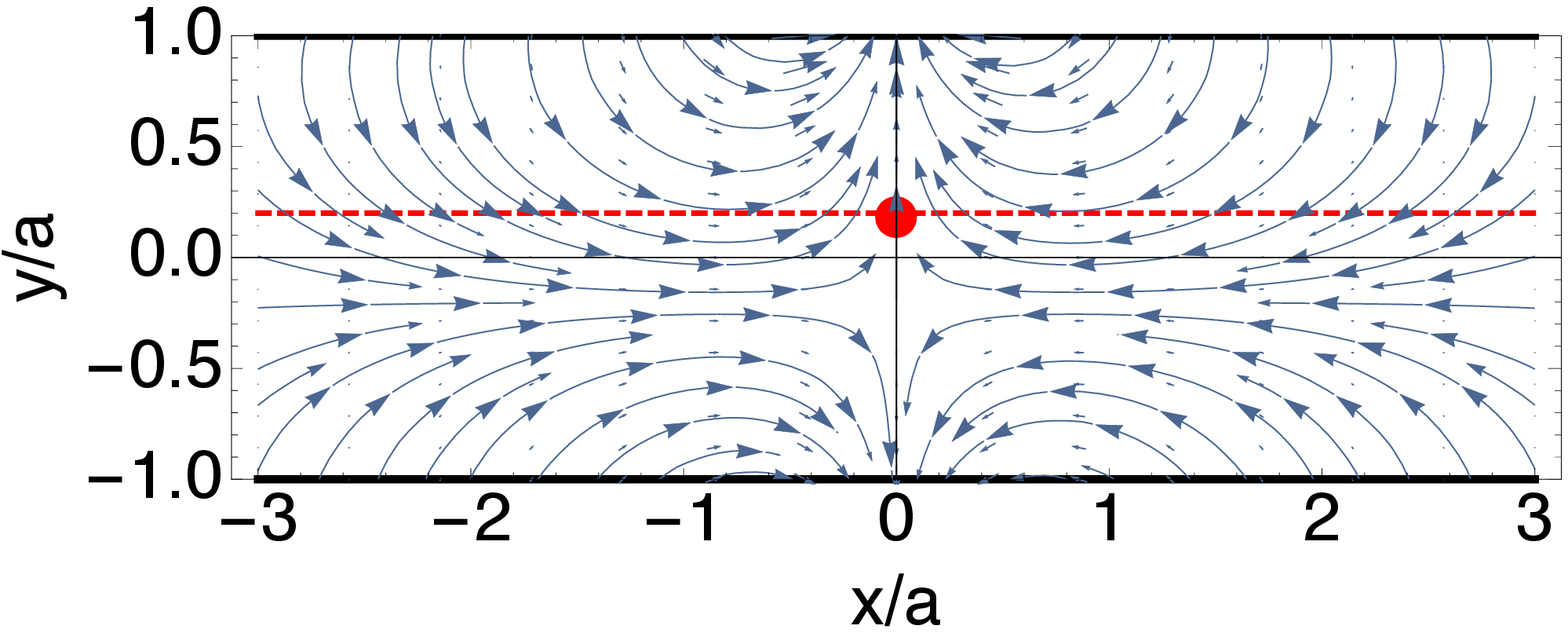}
\caption{Field lines of the transverse wake potential that is given by Eq.\eqref{eq:wp}} for the source (red circle) located in the center of the wakefield structure (left panel) and displaced from the center (right panel) $y_0=0.2a$.
\label{Fig:2m}
\end{figure*}

Now let us consider the cross-section of a planar structure (Fig.\ref{Fig:1} first left panel). 
First we introduce a change of the coordinates to rotate the strip by $\pi/2$ in angle, $\omega(\psi)=i\psi^*$. Then we build a conformal map of the strip onto a circle, as diagrammed in Fig.\ref{Fig:1}.  The plane $\omega$ is mapped onto a circle of radius $a$ with the function 
\begin{align}
\label{eq:m1}
\chi(\omega)=a \tan\left(\frac{\pi}{4}\frac{\omega}{a} \right). 
\end{align}
The point of the bunch location $\omega_0$, is mapped to a point $\chi_0=a \tan\left(\frac{\pi}{4}\frac{\omega_0}{a} \right)$ . 
Then, we map a new circle on this circle such that the point $\chi_0$ corresponds to the center of the last circle. This mapping is accomplished using the function
\begin{align}
\label{eq:m2}
f(\chi,\chi_0)=a^2\frac{\chi-\chi_0}{a^2-\chi\chi_0^*}.
\end{align}  
Using equtions \eqref{eq:Ez} and \eqref{eq:Lort} we can calculate the transverse part of the Lorentz force and longitudinal electric field in $\omega$ plane. 
Combining \eqref{eq:m1} and \eqref{eq:m2} we arrive at
\begin{align}
f(\omega,\omega_0)=a\frac{\tan\left(\frac{\pi}{4}\frac{\omega}{a} \right)-\tan\left(\frac{\pi}{4}\frac{\omega_0}{a} \right)}{1-\tan\left(\frac{\pi}{4}\frac{\omega}{a} \right)\tan\left(\frac{\pi}{4}\frac{\omega_0^*}{a} \right)}.
\end{align}

First we calculate 
\begin{align}
\label{eq:der}
f'(\omega,\omega_0)=\frac{\pi}{4}\frac{\left(\sec\left(\frac{\pi}{4}\frac{\omega}{a} \right)\right)^2\left(1-\left|\tan\left(\frac{\pi}{4}\frac{\omega_0}{a} \right)\right|^2\right)}{\left(1-\tan\left(\frac{\pi}{4}\frac{\omega}{a} \right)\tan\left(\frac{\pi}{4}\frac{\omega_0^*}{a} \right)\right)^2}.
\end{align}
here prime denotes total derivative by $\omega$.

Consequently 
\begin{align}
\label{eq:der0}
f'(\omega_0,\omega_0)=\frac{\pi}{4}\frac{\left(\sec\left(\frac{\pi}{4}\frac{\omega_0}{a} \right)\right)^2}{1-\left|\tan\left(\frac{\pi}{4}\frac{\omega_0}{a} \right)\right|^2}.
\end{align}
Combining \eqref{eq:der} and \eqref{eq:der0} with \eqref{eq:Ez} we have an expression of the longitudinal field in the $\omega$ plane
\begin{align}
\label{eq:Ezo}
E_z^{0^+}(\omega,\omega_0)=-\frac{4Q}{a^2}\frac{\pi^2}{16}\Re\left\{\left[\sec\left(\frac{\pi}{4}\frac{\omega^*+\omega_0}{a} \right) \right]^2\right\}.
\end{align}
Now, we examine the transverse force component and note that 
\begin{align}
\left[F_\bot(\omega,\omega_0)\right]^*=\frac{d}{d\omega} f'(\omega,\omega_0)f'(\omega_0,\omega_0)^*,
\end{align}
and consequently
\begin{align}
\label{eq:Fo}
\left[F_\bot^{\omega}(\omega,\omega_0)\right]^*&=\frac{4 qQ \theta(\zeta) \zeta}{a^3}\frac{\pi^3}{32} \left[\sec\left(\frac{\pi}{4}\frac{\omega+\omega_0^*}{a} \right) \right]^2\times \nonumber \\ &\times\tan\left(\frac{\pi}{4}\frac{\omega+\omega_0^*}{a} \right).
\end{align} 
Making a substitution $\omega(\psi)=i\psi^*$ in \eqref{eq:Ezo} we arrive at the transverse dependence of $E_z$ in $\psi$ plane 
\begin{align}
\label{eq:Ezpl}
E_z^{0^+}(\psi,\psi_0)=-\frac{4Q}{a^2}\frac{\pi^2}{16}\Re\left\{\left[\mathrm{sech}\left(\frac{\pi}{4}\frac{\psi-\psi_0^*}{a} \right) \right]^2\right\}.
\end{align}
Taking into account that $F_\bot^{\psi}=i(F_\bot^{\omega})^*$ with the substitution $\omega(\psi)=i\psi^*$ and \eqref{eq:Fo} we arrive at
 \begin{align}
F_\bot^{\psi}(\psi,\psi_0)&=-\frac{4 qQ \theta(\zeta) \zeta}{a^3}\frac{\pi^3}{32} \left[\mathrm{sech}\left(\frac{\pi}{4}\frac{\psi^*-\psi_0}{a} \right) \right]^2\times \nonumber \\ &\times\tanh\left(\frac{\pi}{4}\frac{\psi^*-\psi_0}{a} \right).
\end{align} 

Since we are only interested in the properties of the distributions,  we introduce scaled wake potentials per unit length in the form
$w_\parallel=-E_z^{0^+}/Q$ and $w_\perp=F^{\psi}_\perp/(qQ)$. With this we finally obtain the longitudinal and transverse wake potentials,
\begin{align}
\label{eq:wl}
w_\parallel(\psi,\psi_0)=\frac{\pi^2}{4a^2}\Re\left\{\left[\mathrm{sech}\left(\frac{\pi}{4}\frac{\psi-\psi_0^*}{a} \right) \right]^2\right\},
\end{align}
 \begin{align}
 \label{eq:wp}
w_\bot(\psi,\psi_0)&=-\frac{\pi^3 \theta(\zeta) \zeta}{8a^3} \left[\mathrm{sech}\left(\frac{\pi}{4}\frac{\psi^*-\psi_0}{a} \right) \right]^2\times \nonumber \\ &\times\tanh\left(\frac{\pi}{4}\frac{\psi^*-\psi_0}{a} \right).
\end{align}

It is noteworthy that the results for the longitudinal and transverse wake potentials for a planar structure in (\ref{eq:wl}) and (\ref{eq:wp}) agree with previously derived results for a rectangular corrugated structure  \cite{BaneDc}.
In Ref.~\cite{BaneDc}, a different approach of surface impedances, developed in \cite{StupSrf}, was employed.   
Both approaches \cite{mySTAB} and \cite{StupSrf,BaneDc} predict the same interesting result, namely, the limiting value of the loss and kick factors are independent of the properties of the retarding material, and therefore the transverse dependencies are properties of the geometry only. For a more detailed explanation of these concepts we refer the reader to the original works \cite{myPRL,mySTAB,StupSrf,BaneDc}.

\section{Mechanism for the transverse wakefield damping}\label{sec:mec}
The effects of the transverse wake potential are manifested in transverse forces that may lead to the growth of the BBU instability. 
There are many methods proposed to stabilize the growth of this effect, such as using external magnetic focusing elements superimposed on the accelerating channel \cite{Qd1,Li:2014,Qd3}  to exploit BNS damping \cite{BNS}. In this section, we explore an alternative method described in \cite{FLb,Mihalcea:2012}, whereby the coupling to the transverse wake is mitigated by employing highly elliptical drive beams in rectangular structures. 

First, we plot field lines for the transverse wake potential \eqref{eq:wp} for the case when a point particle is travelling along the axis of the planar structure (Fig.\ref{Fig:2m} left panel), and second, for an off-axis point particle displaced by $y_0=0.2a$ (Fig.\ref{Fig:2m} right panel). 
From Fig.\ref{Fig:2m} it is apparent that the transverse wake potential has a vortex-like structure away from the origin, 
while near the origin, the field can be described as "quadrupole-like", because the field focuses along $x$-axis and defocuses along $y$-axis.
In the case of the off-axis particle, the vortex-like structure actually leads to focusing in both $y$ and $x$ directions, along the displacement line (dashed line on Fig.\ref{Fig:2m} right panel) for $|x/a|>\sim1$.   

\begin{figure}[t]
\begin{center}
\includegraphics[scale=0.47]{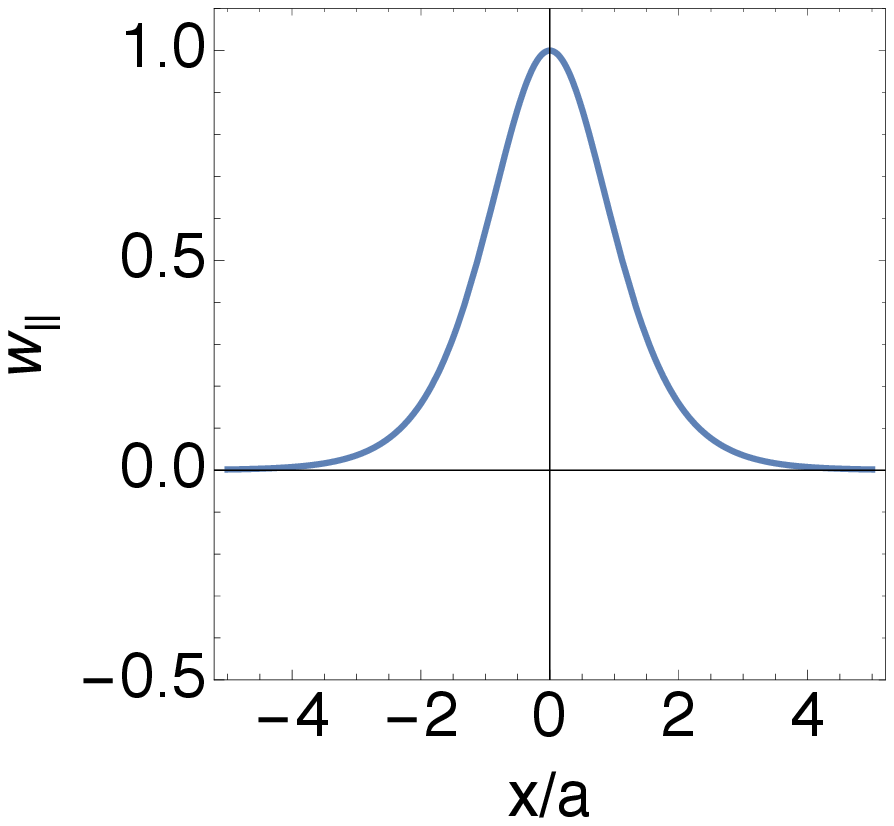}
\includegraphics[scale=0.47]{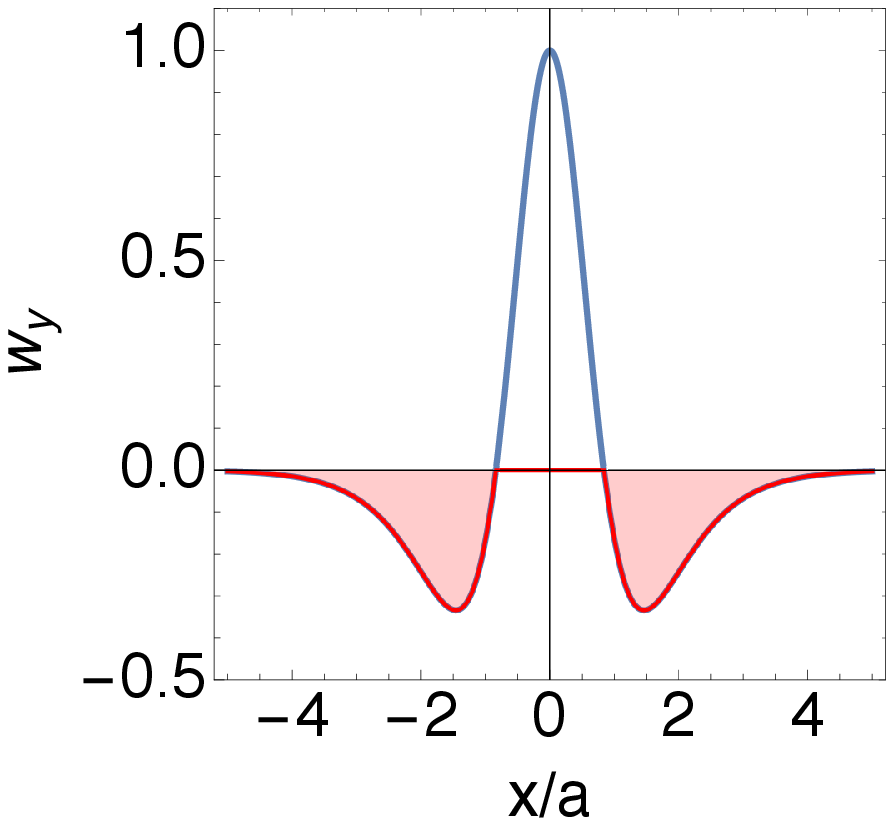}
\caption{$x$-dependence for the longitudinal wake potential (left panel) and $y$-component of the transverse wake potential (right panel) for a point particle. Both plots are normalized to be unity at the point $x=0$. The red shaded area on the right panel shows the focusing region of $w_y$.}
\label{Fig:2}
\end{center}
\end{figure}

For simplicity in further analysis, we assume that the $y$-dimension of the drive bunch is much smaller then the aperture $a$, and we consider small $y_0$ displacements from the center. Using this approximation we decompose \eqref{eq:wl} and \eqref{eq:wp} in a Taylor series at the point $y=y_0=0$ keeping only linear terms in $y$ and $y_0$. Thus, for the longitudinal wake potential we have 
\begin{align}
\label{eq:aplw}
w_\parallel(x,x_0,y,y_0)\approx\frac{\pi^2}{4 a^2}\left[\mathrm{sech}\left(\frac{\pi}{4}\frac{x-x_0}{a}\right)\right]^2,
\end{align} 
and for the $y$-component of the transverse wake potential we have    
\begin{align}
\label{eq:aptw}
&w_y(x,x_0,y,y_0)\approx\frac{\pi^4(y+y_0)\zeta\theta(\zeta)}{32a^4} \times \\ &\times\left[\mathrm{sech}\left(\frac{\pi}{4}\frac{x-x_0}{a}\right)\right]^4\left[2-\cosh\left(\frac{\pi}{2}\frac{x-x_0}{a} \right) \right]. \nonumber
\end{align}  
Formulas \eqref{eq:aplw} and \eqref{eq:aptw} are approximate transverse Green's functions for the longitudinal wake potential and the $y$ component of the transverse wake potential respectively, that are valid for bunches with $\sigma_y<<a$ and small displacements in $y$. 

Now, we assume a transverse bunch distribution of the form
\begin{align}
\rho(x_0,y_0)=\rho_x(x_0)\rho_y(y_0)=\rho_x(x_0)\delta(y_0-0.01a),
\end{align}
and $y=0.01 a$.

We plot the $x$ dependence of $w_\parallel$ and $w_y$ for the point particle with $\rho_x(x_0)=\delta(x_0)$. 
The peak longitudinal field is accompanied by a peak in the transverse wake at $x$=0, as expected, with a tailing off for higher values of $x$.
However, the $y$-component of the transverse wake potential in Fig.\ref{Fig:2}, $w_y$, has a region where it is negative. This implies that instead of being deflected, particles that are located in this region will be attracted back to the $x$-axis. The length of the defocusing region is $2x_c\approx1.68a$, where $x_c$ is the zero crossing in $x$, and the location of the focusing maximum is $x_m\approx\pm1.46a$, as derived in Appendix A.  There are no such features on the longitudinal wake potential, $w_\parallel$.

The existence of this region in the $x$-dependence of $w_y$ wake potential and absence of this feature in the longitudinal wake potential $w_\parallel$ allows one to benefit from stretching the beam in $x$ direction. Increasing the beam size in $x$ diminishes both $w_\parallel$ and $w_y$, however, due to the narrower peak in the $w_y$ pattern, the reduction for $w_y$ will be more pronounced. Furthermore, partial cancelation of $w_y$ is possible due to the focusing regions.

\begin{figure}[t]
\begin{center}
\includegraphics[scale=0.47]{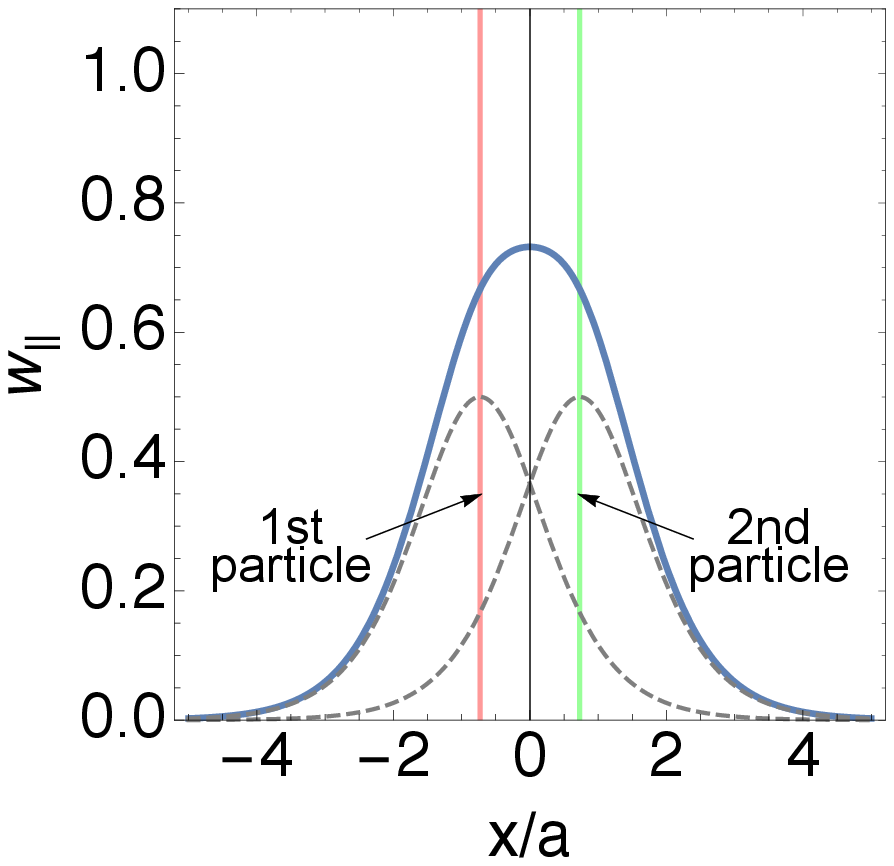}
\includegraphics[scale=0.47]{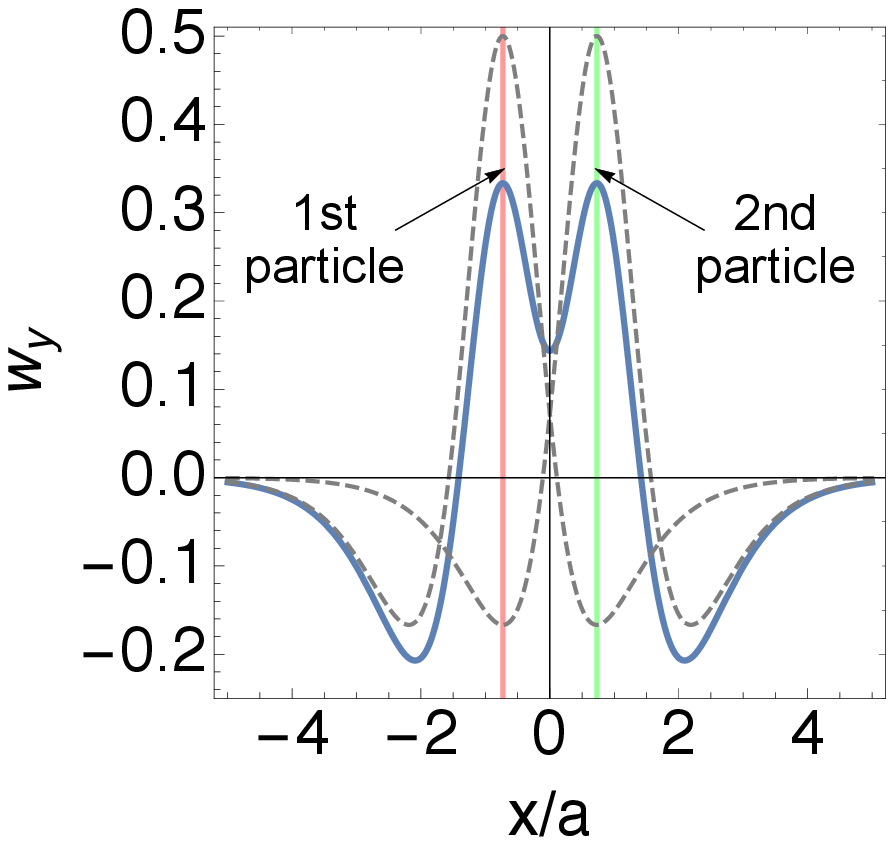}
\caption{$x$-dependence for the longitudinal wake potential (left panel) and $y$ part of the transverse wake potential (right panel) for two point particle separated by a distance equal to the distance to the location of the minimum of the $w_y$. Dashed lines are normalized wake potentials of each particle separately. Solid lines are the resulting wake potentials.}
\label{Fig:3}
\end{center}
\end{figure}

In order to illustrate this feature, we consider a simple example of two synchronous drive particles placed at a distance $|x_m|\approx 1.46 a$ (we assume $\rho(x_0)=1/2[\delta(x_0-0.73a)+\delta(x_0+0.73a)]$), equal to the distance from the origin to the minimum of $w_y$ (See Appendix A).  
This selected arrangement minimizes the amplitude of the combined transverse wake potential of the two particles.
From Fig.\ref{Fig:3} we see that due to decoherence, the longitudinal wake amplitude was reduced by only $27\%$, whereas the transverse wake potential acting on each bunch was reduced by $67\%$ due to the interference effect. Moreover, it is worth mentioning that if a witness particle will be placed at the point $x=0$ after the two driver bunches considered in this example, the maximum kick that the witness bunch will experience is reduced by $86 \%$ compared to standard scheme when a single driver with the same total charge is placed at $x=0$, as seen in Fig.\ref{Fig:3}.

This simple illustration unveils the mechanism behind the idea of transverse wakefield suppression by a highly elliptical drive bunch that was introduced earlier in literature \cite{FLb}.   
In particular, the coherence length in the $x$ direction is greater for the longitudinal wake potential then for the $y$-components of the transverse wake potential. In addition, partial cancelation is possible for the $y$-component of the transverse wake potential due to the existence of the focusing regions in the $x$-dependence of the Green's function $w_y$.

We would like to stress that the analysis above is purely dependent on structure and beam geometry, and independent of the material in the structure, thus is valid for dielectric-lined structures, metallic corrugated structures, resistive wall and the recently considered photonic planar structures \cite{Andonian:2014,Hoang:2018}.

\section{Tradeoff curves}\label{sec:trade}
In this section we further analyze the tradeoff between the loss of the amplitude of the longitudinal wake potential compared to the suppression of the transverse wake potential for varying beam distributions. 
We consider a Gaussian distribution in $x$ and calculate both longitudinal and transverse wake potentials using approximate equations \eqref{eq:aplw} and \eqref{eq:aptw} as
\begin{align}
\label{eq:Wf}
&W_\parallel(x)=\frac{\pi^2}{4 a^2\sigma_x\sqrt{2\pi}}\times  \\ \nonumber &\times\int\limits_{-\infty}^{\infty}\exp\left(-\frac{x_0^2}{2 \sigma_x^2}\right)\left[\mathrm{sech}\left(\frac{\pi}{4}\frac{x-x_0}{a}\right)\right]^2dx_0,
\end{align}
\begin{align}
\label{eq:Wf2}
&W_y(x)=\frac{\pi^4(y+y_0)\zeta\theta(\zeta)}{32a^4\sigma_x\sqrt{2\pi}}\int\limits_{-\infty}^{\infty}dx_0\exp\left(-\frac{x_0^2}{2 \sigma_x^2}\right)\times  \\ \nonumber &\times\left[\mathrm{sech}\left(\frac{\pi}{4}\frac{x-x_0}{a}\right)\right]^4\left[2-\cosh\left(\frac{\pi}{2}\frac{x-x_0}{a} \right) \right].
\end{align}

\begin{figure}[t]
\begin{center}
\includegraphics[scale=0.55]{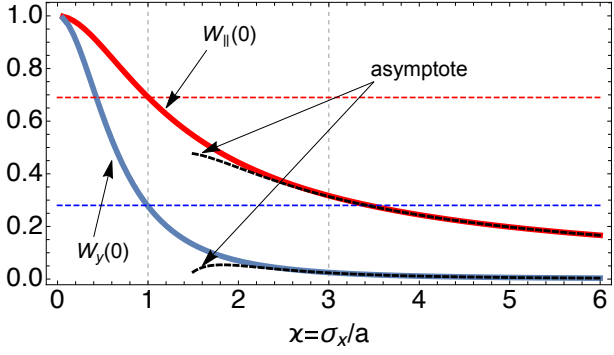}
\caption{Normalized maximum of the longitudinal wake potential (red) and maximum of the $y$-component of the transverse wake potential (blue) for the Gaussian distribution as a functions of the $\sigma_x/a$. Dashed lines are asymptote given by equations \eqref{eq:Wzt} and \eqref{eq:Wyt} }
\label{Fig:4}
\end{center}
\end{figure}

In Fig.\ref{Fig:4} we plot the normalized wake potentials $W_\parallel(0)$ and $W_y(0)$ as a function of the bunch flatness, which we define as  $\varkappa=\sigma_x/a$. When the bunch flatness is increased both longitudinal and transverse wake potentials are reduced. However, the rate of decrease for the transverse wake potential is significantly greater compared to the longitudinal component, as also predicted in Refs. \cite{FLb,Mihalcea:2012}.

To introduce a figure of merit of how fast both wake potentials decrease we calculate asymptotes for relatively large bunch flatnesses ($\sigma_x/a\geq3$) for amplitudes of both longitudinal and transverse wake potentials as a function of $\varkappa=\sigma_x/a$ 
\begin{align}
\label{eq:Wzt}
W_\parallel^{\sigma_x}\approx\frac{4}{a^2}\frac{\sqrt{\pi}}{2\sqrt{2}}\left(\frac{1}{\varkappa}-\frac{2}{3\varkappa^3} \right),
\end{align} 
\begin{align}
\label{eq:Wyt}
W_y^{\sigma_x}\approx\frac{8 (y+y_0)\zeta \theta(\zeta)}{a^4}\frac{\sqrt{\pi}}{4\sqrt{2}}\left(\frac{1}{\varkappa^3}-\frac{2}{\varkappa^5} \right).
\end{align}

 For convenience we introduced the notations $W_\parallel^{\sigma_x}\equiv W_\parallel(0)$, $W_y^{\sigma_x}\equiv W_y(0)$. Derivation of these expressions is found in Appendix \ref{sec:app2} and Appendix \ref{sec:app3} respectively.

The first terms in the brackets of Equation \eqref{eq:Wzt} reduces as $\sim1/\varkappa\sim1/\sigma_x $, while the leading terms in the brackets of Equation~\eqref{eq:Wyt} reduces as $\sim 1/\sigma_x^3$. This scaling demonstrates the favorable tradeoff in longitudinal to transverse wake effects for flat bunches, and is consistent with the predictions in \cite{FLb}.

Until this point, we have only considered the $w_y$ component of the transverse potential. For completeness, we derive the $w_x$ component of the wake potential, taking equation \eqref{eq:wp} and decomposing the real part in a Taylor series at the point $y = y_0 = 0$ keeping only terms linear in $y$ and $y_0$ and arrive at,
\begin{align}
\label{eq:wx}
w_x(x,x_0,y,y_0)\approx-\frac{\pi^3\zeta \theta(\zeta)}{8a^3}\frac{\sinh\left(\frac{\pi}{4}\frac{x-x_0}{a} \right)}{\left[\mathrm{cosh}\left(\frac{\pi}{4}\frac{x-x_0}{a}\right)\right]^3}.
\end{align}

It is immediately apparent that the $x$ focusing force is independent of the source particle transverse displacement in $y_0$ and the $y$ witness particle transverse displacement.
This implies that the effect is present even for a perfectly aligned beam on-axis. 

 Next, we consider a bunch with Gaussian distribution in $x$ and calculate $W_x(x)$ inside this bunch as 
\begin{align}
\label{eq:Wfx}
W_x(x)=-\frac{\pi^3\zeta \theta(\zeta)}{8a^3\sqrt{2\pi}\sigma_x}\int\limits_{-\infty}^{\infty}\frac{dx_0\exp\left(-\frac{x_0^2}{2 \sigma_x^2}\right)\sinh\left(\frac{\pi}{4}\frac{x-x_0}{a} \right)}{\left[\mathrm{cosh}\left(\frac{\pi}{4}\frac{x-x_0}{a}\right)\right]^3}.
\end{align}
As it is shown in Appendix \ref{sec:app4} for the case when $\varkappa=\sigma_x/a>3$ and for $x=\sigma_x$ (the point where focusing force is maximal within the bunch core $x\in[-\sigma_x,\sigma_x]$) $W_x^{\sigma_x}\equiv |W_x(\sigma_x)|$ is given by
\begin{align}
\label{eq:Wxt}
W_x^{\sigma_x}\approx\frac{8\zeta \theta(\zeta)}{a^3}{\frac{\sqrt{\pi}}{4\sqrt{2e}}}\left(\frac{1}{\varkappa^2}-\frac{4}{3\varkappa^4} \right).
\end{align}

\begin{figure}[t]
\begin{center}
\includegraphics[scale=0.55]{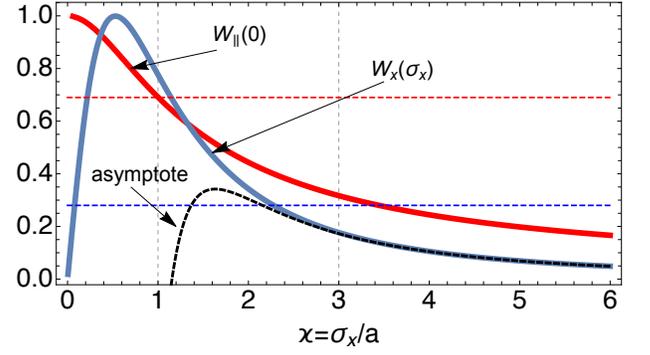}
\caption{Normalized maximum of the longitudinal wake potential (red) and $x$-component of the transverse wake potential at $x=\sigma_x$ (blue) for the Gaussian distribution as a functions of the $\sigma_x/a$. Dashed line is the asymptote given by equation \eqref{eq:Wxt} and multiplied by the same normalization coefficient as blue $x$-component of the transverse wake potential at $x=\sigma_x$.}
\label{Fig:6}
\end{center}
\end{figure}

On Fig.\ref{Fig:6} we plot normalized wakefields $W_\parallel(0)$ and $W_x(\sigma_x)$ as a function of the bunch flatness. One can see that in contrast to the case of $W_y$ when bunch flatness increased $W_x$ first increases up to $\varkappa\sim0.6$ and only then goes down. We also notice that decrease rate for $W_x$ is higher then for $W_\parallel$ but lower then for $W_y$. 

\section{Comparison to cylindrical geometry: Scaled aperture}
Although the mitigating effects of the flat bunch are relatively favorable in regards to optimizing the ratio of the longitudinal to transverse potentials, we must address the fact that there is a reduction to the longitudinal (accelerating) field. In this section, we compare the results derived for the planar structure to the wake potential of a cylindrical structure for a given, i.e. fixed, accelerating gradient. 
In this case, we keep the drive bunch charge fixed and scale the aperture of the planar structure such that the limiting gradient matches that of a cylindrical structure. The maximum gradient in the cylindrical structure per unit charge per unit length for a point particle is given by 
\begin{align}
\label{eq:wc}   
W_\parallel^{c}=\frac{4}{a_c^2}.
\end{align}

Here $a_c$ is the radius of the cylindrical structure.
We equate the gradient in a flat structure \eqref{eq:Wzt} to the gradient in cylindrical structure \eqref{eq:wc} $W_\parallel^{\sigma_x}=W_\parallel^c$ and derive formula for the scaled aperture of the planar structure $a$ to achieve the same gradient as in the cylindrical:
\begin{align}
\label{eq:apt}
a=a_c\frac{\pi^{1/4}}{8^{1/4}}\left(\frac{1}{\varkappa}-\frac{2}{3\varkappa^3} \right)^{1/2}.
\end{align}
Equation (\ref{eq:apt}) gives the scaling law for the aperture in the planar structure and is valid for the $\varkappa=a/\sigma_x\geq3$. Next we substitute \eqref{eq:apt} into the Eq. \eqref{eq:Wyt}, set $y=y_0$, notice that deflecting transverse wakefield for a point particle in the cylindrical structure is given by 
\begin{align}
\label{eq:wtc}   
W_y^{c}=\frac{8y_0\zeta \theta(\zeta)}{a_c^4}
\end{align} 
and arrive at
\begin{align}
\label{eq:trf}
\frac{W_y^{\sigma_x}}{W_y^c}=\sqrt{\frac{8}{\pi}}\frac{9\varkappa^3-18\varkappa}{(3\varkappa^2-2)^2}.
\end{align}
Formula \eqref{eq:trf} shows the reduction in transverse field in a flat structure with flat beam in comparison to cylindrical structure while maintaining the same accelerating gradient. Ultimately for a very flat beam $\varkappa\geq6$, this ratio reduces further to the simple form
 \begin{align}
\label{eq:trf}
\frac{W_y^{\sigma_x}}{W_y^c}\approx\sqrt{\frac{8}{\pi}}\frac{1}{\varkappa}.
\end{align} 
The relation above shows that one can reduce the deflecting wakefield while maintaining an equivalent gradient, by implementing an elliptical beam in a planar structure. The reduction in amplitude, however, is quite modest, scaling as $\sim a/\sigma_x$.      

Now we consider the $W_x$ component of the wake potential that is a property of planar structure and compare the amplitude of $W_x$ to the amplitude of the deflecting wake potential $W_y^c$ in a cylindrical structure.
With equations \eqref{eq:apt} and \eqref{eq:wtc} we express $W_x^{\sigma_x}/W_y^c$ as
\begin{align}
\frac{W_x^{\sigma_x}}{W_y^c}=\frac{1}{(2\pi)^{1/4}}\frac{a_c}{y_0}\sqrt{\frac{3\varkappa}{e}}\frac{3\varkappa^2-4}{(3\varkappa^2-2)^{3/2}}.
\end{align}  
Ultimately for large values of $\varkappa=\sigma_x/a>>1$ we have
\begin{align}
\label{eq:ulim}
\frac{W_x^{\sigma_x}}{W_y^c}\approx \frac{1}{(2\pi)^{1/4} \sqrt{e}}\frac{1}{\sqrt{\varkappa}}\frac{a_c}{y_0}.
\end{align}

From Eq.\ref{eq:ulim}, we can set a conditional statement such that $W_x^{\sigma_x}$ for a planar structure be less then $W_y^c$ of a cylindrical, in the following inequality 
\begin{align}
\varkappa\geq\frac{1}{e\sqrt{2\pi}}\left(\frac{a_c}{y_0}\right)^2.
\end{align}  
For an offset in a cylindrical structure of $a_c/y_0 \sim 10$ we immediately achieve $\varkappa \geq 14.7$. This analysis leads to a conclusion that the aperture scaling approach demands a high degree of bunch flatness to achieve both $y$ and $x$ component amplitudes lower, compared to the deflecting wake in the cylindrical structure.

Indeed, it has been experimentally demonstrated that it is possible to produce beams with very high transverse emittance ratios $\epsilon_x/\epsilon_y=100$  \cite{Piot1} and thus have a high ellipticity $\sigma_x/\sigma_y\sim10$, with more recent work demonstrating emittance ratios $\epsilon_x/\epsilon_y\sim400$ and ellipticity $\sigma_x/\sigma_y>20$ \cite{Piot2,Piot2c}.

It is worth emphasizing this result: we have explicitly shown theoretically that both the $x$ and $y$ components of the transverse wake potential in a planar structure could be simultaneously reduced with the aperture scaling strategy, and both $x$ and $y$ components could be simultaneously lower then the deflecting wake potential in a cylindrical structure with the same amplitude of the longitudinal wake potential.

\section{Comparison to cylindrical geometry: Scaled charge }\label{sec:cs}
The second approach to match the gradient in a planar structure is to increase the charge of the bunch while maintaining the same aperture. 

We start from  equation \eqref{eq:Wzt} for the amplitude of the longitudinal wake potential and compare it to the expression for the cylindrical structure \eqref{eq:wc}. We notice that for the longitudinal electric fields (gradients) to be equal, the charge in the planar structure should be
\begin{align}
\label{eq:char}
Q_{pl}=Q_c\frac{2\sqrt{2}}{\sqrt{\pi}}\frac{3\varkappa^3}{3\varkappa^2-2}.
\end{align}  
Since the dependence on  charge is linear, one can achieve the ratios
\begin{align}
\label{eq:Wp0}
\frac{W_y^{\sigma_x}}{W_y^c}=\frac{3}{\varkappa^2}-\frac{6}{3\varkappa^2-2},
\end{align}
and
\begin{align}
\label{eq:Wxx0}
\frac{W_x^{\sigma_x}}{W_y^c}=\frac{1}{2\varkappa \sqrt{e}}\frac{a}{y_0}\frac{3\varkappa^2-4}{3\varkappa^2-2}.
\end{align}
By simply increasing $W_y^{\sigma_x}$ and $W_x^{\sigma_x}$ by a factor of $\frac{2\sqrt{2}}{\sqrt{\pi}}\frac{3\varkappa^3}{3\varkappa^2-2}$,
taking the limit of large $\varkappa>>1$ we arrive at
\begin{align}
\label{eq:Wp}
\frac{W_y^{\sigma_x}}{W_y^c}\approx\frac{1}{\varkappa^2},
\end{align}
and
\begin{align}
\label{eq:Wxx}
\frac{W_x^{\sigma_x}}{W_y^c}\approx\frac{1}{2\varkappa \sqrt{e}}\frac{a}{y_0}.
\end{align}
From Eqs. \eqref{eq:Wp} and \eqref{eq:Wxx} we see that in case of the charge scaling strategy the decrease in the transverse wake potential is more pronounced than in case of the aperture scaling approach. 
As in the previous section we can set a conditional statement such that $W_x^{\sigma_x}$ for a planar structure be less then $W_y^c$ of a cylindrical. This leads to the following inequality
\begin{align}
\varkappa\geq\frac{1}{2\sqrt{e}}\frac{a_c}{y_0}.
\end{align}  
We repeat the same estimating procedure as in the previous section, from Eq. \eqref{eq:Wxx0} for an offset in a cylindrical structure $a/y_0\sim 10$ and we achieve $\varkappa > 3$ for the $W_x^{\sigma_x}$ to be the lesser then $W_y^c$. 
This value for beam ellipticity is within experimental capabilities for practically achievable flatness, and could indeed be greater, thus further suppressing $W_x$ lesser than the deflecting force in cylindrical structures.

\section{Conclusions}\label{sec:concl} 
In this paper, we have presented an analysis for a planar symmetry slow-wave structure for varying transverse beam flatness $\varkappa=\sigma_x/a$. 
The approach is distinct from previous approaches \cite{FLb,Park} in its generality, derived on the basis of a Green's function approach that accommodates any impedance. In particular, the analysis is independent of the properties of the retarding material used to describe the transverse structure of the wake potential. 
We have explicitly shown that for highly elliptical beams, the transverse wake potential can be dramatically suppressed, and the tradeoff in longitudinal field loss is still favorable for beam ratios greater than $\sigma_x/a\approx$3. Further, we compared the results explicitly to achieve equal accelerating gradients in cylindrical structures by scaling both the aperture and the charge, and propose that charge scaling is a more favorable method. 

As a relevant example, we consider a beam of a total charge $Q=3$nC in a planar structure with a vacuum gap of  $2a=300 \mu$m, comparable to the parameters of the cylindrical structure in Ref.\cite{Oshea:2016}. We consider beam transverse emittance ratios  of $\epsilon_x/\epsilon_y=100$ \cite{Piot1}, and a beta function ratio of $\beta_x/\beta_y=3$ with $\sigma_z=\sigma_y=30\mu$m. From equation \eqref{eq:Wzt} with $\mathrm{max}|E_z|=QW_\parallel^{\sigma_x}$, it is immediately apparent that such a beam can produce longitudinal fields of $\sim$1GV/m in the considered structure. 
In particular, the ultimate limit for the longitudinal electric field is $\mathrm{max}|E_z|\leq 0.83$ GV/m. 
Accordingly, due to the implementation of the flat beam driver, the limiting value for the deflecting Lorentz force drops by almost two orders of magnitude ($\sim$57 times) from $\mathrm{max}|F_y/y_0|\leq 9.7$MV/m/$\mu$m for a round beam with $\sigma_x=\sigma_y$ to $\mathrm{max}|F_y/y_0|\leq 0.17$MV/m/$\mu$m for an elliptical beam with $\sigma_x\sim17.32, \sigma_y\approx520\mu$m and corresponding $\varkappa=\sigma_x/a=3.46$.

The implications of this work are important for investigating possible designs of future wakefield accelerators to combat deleterious transverse fields effects that compromise drive beam stability. The results are general, depending only on geometric factors, and hold true for corrugated, dielectric planar structures and planar structures with resistive walls, and are directly extendable to 1D and 3D photonic-like planar structures that allow for further precision modal control when driven by selectively shaped beams \cite{Andonian:2014,Hoang:2018}.  

\begin{acknowledgments}
This work was supported by the U.S. National Science Foundation under Award No. PHY-1549132, the Center for Bright Beams and under Award No. PHY-1535639; by the U.S. Department of Energy Award No. DE-SC0017648 and by the U.S. Department of Energy Award No. DE-SC0009914.
The authors are grateful to Alexander Zholents for useful discussions and suggestions.      
\end{acknowledgments}

\appendix
\section{\label{sec:app1} Derivation of the location of the zeros $x_c$ and coordinate of the minimum $x_m$ of the function $w_y$}
We assume a point particle to be placed at the $x_0=0$, in this case equation \eqref{eq:aptw} reads
\begin{align}
\label{eq:aptw1}
&w_y(x,0,y,y_0)\approx\frac{\pi^4(y+y_0)\zeta\theta(\zeta)}{32a^4} \times \\ &\times\left[\mathrm{sech}\left(\frac{\pi}{4}\frac{x}{a}\right)\right]^4\left[2-\cosh\left(\frac{\pi}{2}\frac{x}{a} \right) \right]. \nonumber
\end{align}  
Equating wake potential to zero $w_y(x,0,y,y_0)=0$ we arrive to the following equation on $x_c$ (zeros of wake potential by $x$)

\begin{align}
\label{eq:zeq}
2-\cosh\left(\frac{\pi}{2}\frac{x_c}{a} \right)=0.
\end{align}
Expanding hyperbolic cosine using well known formula 
\begin{align}
\label{eq:cex}
\cosh(s)=\frac{e^s+e^{-s}}{2}
\end{align}
and introducing notation 
\begin{align}
\label{eq:Xdef}
X\equiv \exp \left[\frac{\pi}{2}\frac{x_c}{a} \right],   
\end{align}
we rewrite \eqref{eq:zeq} as
\begin{align}
X_c^2-4X_c+1=0.
\end{align}
Solution of this equation is
\begin{align}
X_c=2\pm\sqrt{3}.
\end{align}
Consequently with substitution \eqref{eq:Xdef} we have for the zeros of $w_y$
\begin{align}
x_c=a\frac{2}{\pi}\log\left[2\pm\sqrt{3} \right].
\end{align}
Next let us find location of the minimums $x_m$ of the function $w_y$. Following standard procedure we equate $x$-derivative of equation \eqref{eq:aptw1} to zero $\frac{\partial w_y}{\partial x}=0$ and arrive at
\begin{align}  
\left[\cosh \left(\frac{\pi}{2}\frac{x}{a}\right)-5\right] \tanh \left(\frac{\pi}{4}\frac{x}{a}\right) \left[\mathrm{sech}\left(\frac{\pi}{4}\frac{x}{a}\right)\right]^4=0.
\end{align} 
As far as $x=0$ is the coordinate of maximum for the coordinates of minimums we have an equation
\begin{align}
\cosh \left(\frac{\pi }{2}\frac{x}{a}\right)-5=0.
\end{align} 
We once again expanding hyperbolic cosine as \eqref{eq:cex} and introducing substitution \eqref{eq:Xdef} we arrive at the algebraic equation
\begin{align}
X_m^2-10X_m+1=0.
\end{align}
Solution of this equation is
\begin{align}
X_m=5\pm2\sqrt{6}.
\end{align}
Consequently with substitution \eqref{eq:Xdef} we have for the positions of the minimums of $w_y$
\begin{align}
x_m=a\frac{2}{\pi}\log\left[5\pm2\sqrt{6} \right].
\end{align}

\section{\label{sec:app2} Asymptote for the maximal amplitude of the longitudinal wake potential $W_{\parallel}(0)$ for large ratios $\sigma_x/a>3$}
First we consider equation for the longitudinal as given by Eq.\eqref{eq:Wf} and rewrite it in the following form for the point $x=0$, where function $W_\parallel(x)$ has maximum
\begin{align}
\label{eq:kp}
W_\parallel^{\sigma_x}=\frac{\pi^{3/2}}{4a^2\sqrt{2}\varkappa}\int\limits_{-\infty}^{\infty}\exp\left(-\frac{\tilde{x}^2}{2 \varkappa^2}\right)\left[\mathrm{sech}\left(\frac{\pi}{4}\tilde{x}\right)\right]^2d\tilde{x}
\end{align}
with $\varkappa=\sigma_x/a$. 

Assuming $1/\varkappa<<1$ we decompose exponent under the integral in Taylor series and keep only first two terms
\begin{align}
\label{eq:expt}
\exp\left(-\frac{\tilde{x}^2}{2 \varkappa^2}\right)\approx 1 - \frac{\tilde{x}^2}{2 \varkappa^2}.
\end{align} 
After the substitution of \eqref{eq:expt} into \eqref{eq:kp} we have
\begin{align}
\label{eq:kp2}
W_\parallel^{\sigma_x}\approx\frac{\pi^{3/2}}{4a^2\sqrt{2}}\left(\frac{I_1}{\varkappa}-\frac{I_2}{2\varkappa^3} \right),
\end{align}
with
\begin{align}
I_1&=\int\limits_{-\infty}^{\infty}\left[\mathrm{sech}\left(\frac{\pi}{4}\tilde{x}\right)\right]^2d\tilde{x}, \\
I_2&=\int\limits_{-\infty}^{\infty}\tilde{x}^2\left[\mathrm{sech}\left(\frac{\pi}{4}\tilde{x}\right)\right]^2d\tilde{x}.
\end{align}   
After substitution of variables and integrating by parts in case of the second integral $I_2$ one may show that
\begin{align}
\label{eq:intv}
I_1=\frac{8}{\pi}, \nonumber \\
I_2=\frac{32}{3\pi}.
\end{align}
We combine \eqref{eq:kp2} and \eqref{eq:intv} and arrive at the final formula for the asymptote
\begin{align}
\label{eq:kpf}
W_\parallel^{\sigma_x}\approx\frac{4}{a^2}\frac{\sqrt{\pi}}{2\sqrt{2}}\left(\frac{1}{\varkappa}-\frac{2}{3\varkappa^3} \right).
\end{align}

\section{\label{sec:app3} Asymptote for the maximal amplitude of the y-component of the transverse wake potential $W_{y}(0)$ for the large ratios $\sigma_x/a>3$}
As in the Appendix \ref{sec:app2} we rewrite amplitude $\kappa_y^{\sigma_x}$ given by
\begin{align}
\kappa_y^{\sigma_x}\equiv\frac{W_y(0)}{(y+y_0)\zeta \theta(\zeta)}
\end{align}
for the $y$-component of the transverse wake potential for a Gaussian bunch \eqref{eq:Wf2}  in the form
\begin{align}
\label{eq:kt}
&\kappa_y^{\sigma_x}=\frac{\pi^{7/2}}{32\sqrt{2}a^4\varkappa}\int\limits_{-\infty}^{\infty} d\tilde{x}\exp\left(-\frac{\tilde{x}^2}{2 \varkappa^2}\right)\times \\ \nonumber
&\times\left[\mathrm{sech}\left(\frac{\pi}{4}\tilde{x}\right)\right]^4\left[2-\cosh\left(\frac{\pi}{2}\tilde{x} \right) \right].
\end{align} 
As before we assume $1/\varkappa<<1$ and decompose exponent under the integral in Taylor series. Now we keep first three terms
\begin{align}
\label{eq:expt2}
\exp\left(-\frac{\tilde{x}^2}{2 \varkappa^2}\right)\approx 1 - \frac{\tilde{x}^2}{2 \varkappa^2}+\frac{\tilde{x}^4}{8 \varkappa^4}.
\end{align} 
After the substitution of \eqref{eq:expt2} into \eqref{eq:kt} we have
\begin{align}
\label{eq:kt21}
&\kappa_y^{\sigma_x}\approx\frac{\pi^{7/2}}{32\sqrt{2}a^4}\left(\frac{I_3}{\varkappa}-\frac{I_4}{2 \varkappa^3}+\frac{I_5}{8\varkappa^5}\right)
\end{align}
with
\begin{align}
I_3&=\int\limits_{-\infty}^{\infty}\left[\mathrm{sech}\left(\frac{\pi}{4}\tilde{x}\right)\right]^4\left[2-\cosh\left(\frac{\pi}{2}\tilde{x} \right) \right]d\tilde{x}, \\
I_4&=\int\limits_{-\infty}^{\infty}\tilde{x}^2\left[\mathrm{sech}\left(\frac{\pi}{4}\tilde{x}\right)\right]^4\left[2-\cosh\left(\frac{\pi}{2}\tilde{x} \right) \right]d\tilde{x}, \\
I_5&=\int\limits_{-\infty}^{\infty}\tilde{x}^4\left[\mathrm{sech}\left(\frac{\pi}{4}\tilde{x}\right)\right]^4\left[2-\cosh\left(\frac{\pi}{2}\tilde{x} \right) \right]d\tilde{x}.
\end{align}   
After substitution of variables and integrating by parts one may show that
\begin{align}
\label{eq:iv2}
I_3&=0, \nonumber \\
I_4&=-\frac{128}{\pi^3}, \\
I_5&=-\frac{1024}{\pi^3} \nonumber.
\end{align}
We combine \eqref{eq:kt21} and \eqref{eq:iv2} and arrive at the final formula for the asymptote
\begin{align}
\label{eq:kpf}
\kappa_y^{\sigma_x}\approx\frac{8}{a^4}\frac{\sqrt{\pi}}{4\sqrt{2}}\left(\frac{1}{\varkappa^3}-\frac{2}{\varkappa^5} \right).
\end{align}  

\section{\label{sec:app4} Asymptote for the maximal amplitude of the x-component of the transverse wake potential $W_{x}(\sigma_x)$ for the large ratios $\sigma_x/a>3$}
We introduce amplitude $\kappa_x^{\sigma_x}$ as
\begin{align}
\kappa_x^{\sigma_x}\equiv\frac{|W_x(\sigma_x)|}{\zeta \theta(\zeta)}
\end{align}
for the $x$-component of the transverse wake potential for a Gaussian bunch \eqref{eq:Wfx}  in the form
\begin{align}
\label{eq:kx}
\kappa_x^{\sigma_x}=\frac{\pi^{5/2}}{8\sqrt{2}a^3\varkappa}\left|\int\limits_{-\infty}^{\infty}\frac{d\hat{x}\exp\left(-\frac{(\hat{x}+\varkappa)^2}{2 \varkappa^2}\right)\sinh\left(\frac{\pi}{4}\hat{x} \right)}{\left[\mathrm{cosh}\left(\frac{\pi}{4}\hat{x}\right)\right]^3}\right|.
\end{align} 
We assume $1/\varkappa<<1$ and decompose exponent under the integral in Taylor series and keep first three terms
\begin{align}
\label{eq:expt3}
\exp\left[-\frac{(\hat{x}+\varkappa)^2}{2 \varkappa^2}\right]\approx \frac{1}{\sqrt{e}} - \frac{\hat{x}}{\varkappa\sqrt{e}}+\frac{\hat{x}^3}{3\varkappa^3\sqrt{e}}.
\end{align} 
After the substitution of \eqref{eq:expt3} into \eqref{eq:kx} we have
\begin{align}
\label{eq:kt2}
\kappa_x^{\sigma_x}\approx\frac{\pi^{5/2}}{8\sqrt{2 e}a^3}\left|\frac{I_6}{\varkappa}-\frac{I_7}{\varkappa^2}+\frac{I_8}{3\varkappa^4} \right|
\end{align}
with
\begin{align}
I_6&=\int\limits_{-\infty}^{\infty}\frac{\sinh\left(\frac{\pi}{4}\hat{x} \right)}{\left[\mathrm{cosh}\left(\frac{\pi}{4}\hat{x}\right)\right]^3}d\hat{x}, \\
I_7&=\int\limits_{-\infty}^{\infty}\hat{x}\frac{\sinh\left(\frac{\pi}{4}\hat{x} \right)}{\left[\mathrm{cosh}\left(\frac{\pi}{4}\hat{x}\right)\right]^3}d\hat{x}, \\
I_8&=\int\limits_{-\infty}^{\infty}\hat{x}^3\frac{\sinh\left(\frac{\pi}{4}\hat{x} \right)}{\left[\mathrm{cosh}\left(\frac{\pi}{4}\hat{x}\right)\right]^3}d\hat{x}.
\end{align}   
After substitution of variables and integrating by parts one may show that
\begin{align}
\label{eq:iv3}
I_6&=0, \nonumber \\
I_7&=\frac{16}{\pi^2}, \\
I_8&=\frac{64}{\pi^2} \nonumber.
\end{align}
We combine \eqref{eq:kt2} and \eqref{eq:iv3} and arrive at the final formula for the asymptote
\begin{align}
\label{eq:kpf}
\kappa_x^{\sigma_x}\approx\frac{8}{a^3}{\frac{\sqrt{\pi}} {4\sqrt{2e}}}\left(\frac{1}{\varkappa^2}-\frac{4}{3\varkappa^4} \right).
\end{align}  

\bibliographystyle{ieeetr}
\bibliography{Transverse_shaping}

\end{document}